\documentclass[prb,twocolumn,amssymb,showpacs,superscriptaddress]{revtex4}
\usepackage{amsmath} 
\usepackage{amssymb} 
\usepackage{arydshln}
\usepackage{graphicx, subfigure}
\usepackage{epsfig}
\usepackage{epstopdf}

\usepackage{color}
\usepackage{rotating}
\usepackage{bm}

\begin{document}
\title{Triplet superconductivity in a model of Li$_{0.9}$Mo$_6$O$_{17}$}
\author{Natalia Lera}
\affiliation{CIC nanoGUNE, 20018 Donostia-San Sebasti\'an, Spain}
\affiliation{Departamento de F\'isica de la Materia Condensada, 
Universidad Aut\'onoma de Madrid, Madrid 28049, Spain}
\author{J.V. Alvarez}

\affiliation{Departamento de F\'isica de la Materia Condensada, 
Universidad Aut\'onoma de Madrid, Madrid 28049, Spain}
\affiliation{Condensed Matter Physics Center (IFIMAC)
and Instituto Nicol\'as Cabrera,
Universidad Aut\'onoma de Madrid, Madrid 28049, Spain}
\date{\today}
\begin{abstract}
Superconductivity in the quasi-one-dimensional material Li$_{0.9}$Mo$_6$O$_{17}$ is analyzed based on a multiorbital extended Hubbard model. We found strong charge fluctuations at two different momenta ${\bf Q_1}$ and ${\bf Q_2}$ giving rise to two different charge ordered phases. Evaluating the superconducting vertex, we found superconductivity near strong charge fluctuations at ${\bf Q_1}$.  The order parameter has the p-wave symmetry with nodes on the Fermi surface. The metallic state displays a characteristic charge collective mode ${\bf Q_1}$ due to nesting and for on-site Hubbard repulsion sufficiently large, a charge critical mode ${\bf Q_2}$ driven by Coulomb repulsion, which softens at the proximity to the transition. The results are quite robust for different coupling parametrizations. A phase diagram discussing the relevance of the model to the physics of the material  is proposed. 
\end{abstract}
\pacs{71.10.Hf, 71.10.Fd, 74.40.Kb, 74.70.-b}
\maketitle

\section{Introduction}

Low-temperature physics of correlated materials is often characterized by the competition between ordered phases and unconventional superconductivity. 
Tipically, a static mean-field description, implying negligible fluctuations beyond the limits of the ordered phase, is not valid in these systems. Nearly all dynamical probes show strong order parameter fluctuations, not only in the neighboring superconducting phases, which suggests a natural mechanism of pairing, but also in the strange metal, present at higher temperatures. Lithium Purple Bronze (LiPB), adds the ingredient of quasi-one-dimensionality to the problem and suggests the possibility that charge and spin fluctuations alone, without the existence of  real order, might be responsible of superconductivity and anomalies of the normal phase. 

The metallic phase of LiPB, with chemical formula Li$_{0.9}$Mo$_6$O$_{17}$,  has been characterized as a robust  Luttinger Liquid (LL)  in a series of Angle Resolved Photoemission Spectroscopy (ARPES) experiments ranging different temperature regimes,  sample growth techniques, photon energies, and data analysis procedures   \cite{denlinger,gweon2001,gweon2002,gweon2003,gweon2004,SolidStateAllen,wang2006,wang2009,dudy2013}. STM spectroscopy shows  \cite{cazalilla2005,Matzdorf2013}  LL single-particle density of states  and thermal and electric transport measurements are in complete  disagreement with Widemann-Franz law \cite{hussey2011}.  When temperature is decreased, an upturn of the resistivity occurs at $T_m \sim 20 $ K \cite{Greenblatt,filippini1989,Choi,hussey2011}  and the  material becomes superconducting at lower temperatures around $T_c \sim 1$ K \cite{filippini1989, mercure2012}.  

Unlike other low-dimensional bronzes, the resistivity upturn of  LiPB \cite{dumas1985} is not associated with a lattice distortion (See Table 1 in Ref. \onlinecite{VanderSmaalen}).  Neither thermal expansion \cite{dossantos2007}  nor  neutron scattering experiments \cite{daluz2011}  have identified a phase transition at $T_m$ suggesting the idea of a soft crossover of electronic nature. No gap has been clearly observed in the spectroscopies but optical conductivity measurements \cite{Choi}  suggest the presence of a weak pseudogap. Recently,   thermopower \cite{Cohn} and NMR   \cite{Wu}  experiments have confirmed different aspects of the quasi-one-dimensionality of this material  but  the nature of the upturn remains a mistery.

The most recent study of superconducting properties  \cite{mercure2012}  confirms quantitatively  that the  large anisotropies  observed in the upper critical field agree with those expected from the electrical resistivity in the metallic phase. The coherence lengths perpendicular to the chains are larger 
than interchain distances and $H_{c2}$ increases monotonically with decreasing temperature to values 5 times larger than the estimated paramagnetic pair-breaking field. Neither spin-orbit scattering nor strong-coupling superconductivity seem to explain this behavior suggesting the possibility of spin triplet superconductivity. 
 A quantitative comparison with experiments \cite{lebed}  shows that  superconductivity can be destroyed through
orbital effects at fields higher than the Clogston paramagnetic limit {\em  provided} that the superconducting pairs are in the triplet state.\\

In the last years there has been a very important theoretical effort  \cite{merino2012,chudzinsky2012,JMandJV,nuss}  to reduce the complexity of the unit cell 
to microscopic Hamiltonians reproducing different aspects of this phenomenology. In this article, we present a microscopic theory for the unconventional superconducting properties observed in  
Li$_{0.9}$Mo$_6$O$_{17}$.  Based on a minimal extended Hubbard model introduced in Ref. \onlinecite{merino2012,chudzinsky2012}, 
we show that Li$_{0.9}$Mo$_6$O$_{17}$ superconducts in the triplet channel when charge and spin fluctuations are enhanced,  
which may be also related with the upturn in resistivity at $T_m$ \cite{JMandJV}. Using the random phase approximation (RPA), we identify the CDW pattern characterized by two ordering wave vectors, ${\bf Q_1}$ and ${\bf Q_2}$. In the proximity of those phases we evaluate and analize the superconducting vertex finding dominant p-wave triplet superconductivity with nodes on the Fermi surface.  Within our methodology  we  
find results  compatible with the one presented  in a  very recent preprint \cite{TripletLiPB}

\begin{figure}
\epsfig{file=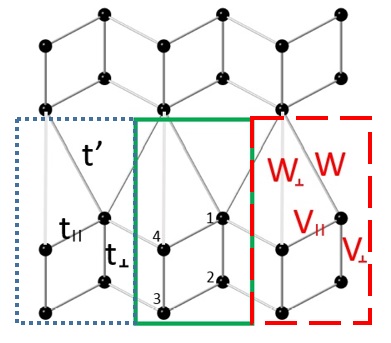,width=8cm}
\caption{(Color online) 
Schematic crystal structure of Li$_{0.9}$Mo$_6$O$_{17}$ projected onto the $b$-$c$ plane showing only the partially filled Mo atoms forming the zig-zag ladders relevant to the low energy electronic properties. Our choice of unit cell is highlighted and the orbitals numerated (solid line) according to the text, the hoppings (dotted line) and the Coulomb interactions (dashed line) are also represented. 
 }
\label{fig1}
\end{figure}

\section{Microscopic model.}
\label{sec:model}

The electronic structure close to the $E_F$ and the quasi-one-dimensionality of the system derives from two parallel zig-zag Mo-O chains per unit cell \cite{onoda1987} Fig. \ref{fig1}. Tight binding \cite{wangbo} and DFT \cite{popovic2006} band structure calculations agree that the Mo-O orbitals of the chain give rise to four bands and two of them cross the Fermi level. 
ARPES confirms the quasi-one-dimensionality of the Fermi surface. A Slater-Koster 
tight binding parametrization of the system has been propoposed in Ref. \onlinecite{merino2012} and the role of long-range Coulomb couplings
in the anomalies of the metallic phase has been also studied \cite{JMandJV} . 
Here, we consider a strongly correlated model, which can capture the essential physics of Li$_{0.9}$Mo$_6$O$_{17}$ \cite{merino2012} consisting on an extended Hubbard lattice with 4 Mo-atoms per unit cell, which reads:
\begin{equation}
H=H_0+H_U,
\label{eq:model}
\end{equation}
where $H_0$ is the non-interacting tight-binding Hamiltonian.
The one-electron Hamiltonian can be expressed in terms of Bloch waves
with the following non-zero matrix elements, the intra-ladder: $t_{12}({\bf k})=t_{43}({\bf k}) =t_\perp=-0.024$ eV,and $t_{14}=t_{23}({\bf k}) t=0.5$ eV and the hoppings among chains: $t_{13}({\bf k})=t'=0.036$ eV, as is shown in Fig \ref{fig1} (dotted cell).

\begin{figure}
  \begin{tabular}{cc}
    \includegraphics[width=40mm,height=40mm]{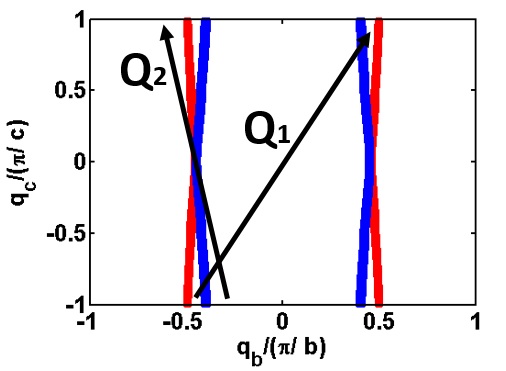}
    \includegraphics[width=50mm,height=40mm]{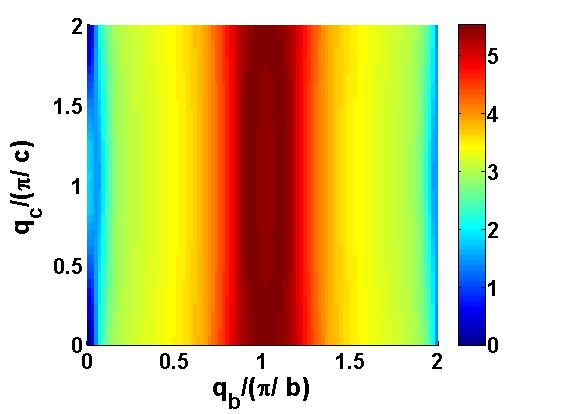}

  \end{tabular}
\caption{(Color online) 
(a) Fermi surface with two bands, ${\bf Q_1}$ a nesting vector, and ${\bf Q_2}$ referred in the text.(b)Real part of the bare susceptibility in momentum space for $\omega=0$ and $q_a=0$. Notice the maximum reveals the warping of the Fermi surface at the nesting vector.}
\label{chi0}
\end{figure}

The diagonalized Hamiltonian: $H_0=\sum_{{\bf k}\mu \sigma} \epsilon_\mu({\bf k} ) d^\dagger_{{\bf k} \mu \sigma} d_{{\bf k} \mu \sigma}$, 
leads to four bands denoted by $\mu$, the two lowest ones cross the $E_F$ \cite{popovic2006,merino2012,JMandJV}. The Fermi surface, 
close to one quarter-filling, $n=0.225$, is shown in Fig. \ref{chi0} (a).

The Coulomb interaction terms in the Hamiltonian 
includes on-site Hubbard interaction ($U$), intra-ladder interaction with the following non-zero matrix elements: $V_{12}=V_{32}=V_{\parallel}$ and $V_{12}=V_{34}=V_{\perp}$ and inter-ladder $W$ interactions: $W_{13}=W$ and $W_{12}=W_{34}=W_\perp$, as is shown in Fig \ref{fig1} (dashed cell).

\begin{multline}
H_U = U\sum_{l,i,\alpha} n^{(l)}_{i \alpha \uparrow} n^{(l)}_{i \alpha\downarrow}
+\sum_{l,i,\alpha,j,\beta}V_{i \alpha,j \beta} n^{(l)\dagger}_{i \alpha} n^{(l)}_{j \beta} 
 \\
+\sum_{l,i, \alpha,j, \beta}W_{i \alpha,j \beta} n^{(l)}_{i \alpha} n^{(l+1)}_{j \beta}
 \\
\label{eq:hu}
\end{multline}

The interacting Hamiltonian only includes density-density Coulomb interaction contributions. 
Within this work, we have consider several combinations of parameters, all of them leading to essentially the same results presented here where we reduce the parameter space to  two variables ($U$ and $V$). We take the Coulomb interaction among different sites with $1/|{\bf r}|$ dependence, where ${\bf |r|}$ is the distance among orbitals. Therefore, we parametrize the interactions by weighting the V's with the interatomic distances: $V=V_\parallel r_\parallel=V_\perp r_\perp=Wr_W=W_\perp r_{W\perp}$. 

\section{Multiorbital RPA approach}
In this section we explain the multi-orbital random phase approximation (RPA) approach for this model, we will study spin and charge ordering based on spin and charge susceptibility respectively, and the superconducting vertex based on projections of different order parameters.
\subsection{Spin susceptibility}
The RPA spin susceptibility reads\cite{maier2009}: 

\begin{multline}
(\chi_s)_{\alpha,\beta} ({\bf q})=(\chi_0)_{\alpha, \beta}({\bf q})\\
+\sum_{\alpha ' \beta '}(\chi_s)_{\alpha ' \beta '}({\bf q})(U_s)^{\alpha ' \beta '}(\chi_0)_{\alpha \beta}({\bf q})
\label{eq:chiS}
\end{multline}
where the indices $\alpha,\beta$ refer to the four Mo $d_{xy}$ orbitals present in the unit cell. This is the more general case for density-density interactions. In our case the spin interaction is a diagonal matrix $(U_s)_{\alpha\beta}=U \delta_{\alpha,\beta}$, momentum independent.
The non-interacting susceptibility, $\chi_0$, reads: 
\begin{widetext}
\begin{equation}
(\chi_0)_{\alpha,\beta}({\bf q}, i\omega)=-{1 \over N} \sum_{\bf k, \mu, \nu} {a^\alpha_\mu({\bf k}) a^{\beta *}_\mu({\bf k})  a^{\beta}_\nu({\bf k + q}) a^{\alpha *}_\nu({\bf k +q}) \over  i\omega +
\epsilon_\nu({\bf k +q}) -\epsilon_\mu({\bf k}) } [ f(\epsilon_\nu({\bf k+q}))-f(\epsilon_\mu({\bf k}) ) ] ,
\label{eq:chi0}
\end{equation}
\end{widetext}
where $N$ is the number of lattice sites, and $\nu,\mu$ are band indices.  The matrix elements $a^\alpha_\mu({\bf k})=\langle \alpha | \mu {\bf k}  \rangle $ are 
the coefficients of the eigenvectors diagonalizing $H_0$.

\subsection{Charge susceptibility}
The RPA charge susceptibility reads\cite{maier2009}: 
\begin{multline}
(\chi_c)_{\alpha,\beta} ({\bf q})=(\chi_0)_{\alpha, \beta}({\bf q})\\
-\sum_{\alpha ' \beta '}(\chi_c)_{\alpha ' \beta '}({\bf q})(U_c)^{\alpha ' \beta '}({\bf q})(\chi_0)_{\alpha \beta}({\bf q})
\label{eq:chiC}
\end{multline}
Where ${U}_c({\bf q})$ is the Coulomb matrix appearing in Eq. (\ref{eq:hu}) expressed in momentum space: $({U}_c)_{\alpha \beta}({\bf q})=U\delta_{\alpha,\beta}+2\hat V({\bf q})_{\alpha,\beta}$
where $\hat V({\bf q})$ is the Fourier transform of $V_{i \alpha,j \beta}$ and $W_{i \alpha,j \beta}$ interactions in real space.
\subsection{Superconducting Vertex}
Assuming that the pairing interaction arises from the exchange of spin and charge fluctuations, we can calculate the
pairing vertex using the RPA, (For a detailed description of the method, see for instance \cite{maier2009}). The strength of the interaction is weighted by $\omega^{-1}$ and making use of the Kramers-Kronig relation we only need the zero frequency vertex,\cite{maier2009}. For the multiorbital case \cite{Takimoto}\cite{japoneses}, singlet and triplet pairing vertex at zero frequency are given by:
\begin{widetext}
\begin{equation}
\Gamma_{\alpha \beta}^{\mathrm{singlet}}({\bf k,k'})= \left ( U+\frac{3}{2}U_s \chi_s({\bf k-k'})U_s+\hat V({\bf k-k'})
-\frac{1}{2}U_c({\bf k-k'}) \chi_c({\bf k-k'})U_c({\bf k-k'}) \right )_{\alpha \beta}
\end{equation}

\begin{equation}
\Gamma_{\alpha \beta}^{\mathrm{triplet}}({\bf k,k'})= \left ( -\frac{1}{2}U_s \chi_s({\bf k-k'})U_s+\hat V({\bf k-k'})
-\frac{1}{2}U_c({\bf k-k'}) \chi_c({\bf k-k'})U_c({\bf k-k'}) \right ) _{\alpha \beta} 
\label{GTriplet}
\end{equation}
\end{widetext}
We transform the vertex in real space $\alpha \beta$ into momentum space $\mu \nu$ with the band structure eigenvalues $a^\alpha_\mu({\bf k})$. The Cooper pairs have an incoming momentum of ($\bf k$,$\bf -k$) and an outcoming momentum of ($\bf k'$,$\bf -k'$). We take the symmetric and antisymmetric parts for singlet and triplet channels respectively.
\begin{widetext}
\begin{equation}
\Gamma_{\mu \nu}^{\mathrm{singlet}}({\bf k,k'})=\sum _{\alpha \beta} 
a^{\alpha*}_\mu({\bf -k})a^{\alpha*}_\mu({\bf k})
\mathrm{Real}\left [ \Gamma_{\alpha \beta}^{\mathrm{singlet}}({\bf k,k'}) \right ]
a^{\beta}_\nu({\bf k'})a^{\beta}_\nu({\bf -k'})+
({\bf k'}\leftrightarrow {\bf -k'})
\label{pairingVertexs}
\end{equation}

\begin{equation}
\Gamma_{\mu \nu}^{\mathrm{triplet}}({\bf k,k'})=\sum _{\alpha \beta} 
a^{\alpha*}_\mu({\bf -k})a^{\alpha*}_\mu({\bf k})
\mathrm{Real}\left [ \Gamma_{\alpha \beta}^{\mathrm{triplet}}({\bf k,k'}) \right ]
a^{\beta}_\nu({\bf k'})a^{\beta}_\nu({\bf -k'})-
({\bf k'}\leftrightarrow {\bf -k'})
\label{pairingVertext}
\end{equation}
\end{widetext}
We solve the gap equation by projecting out $s$,$p$,$d$ and $f$ waves (\cite{Scalapino89}):
\begin{equation}
\lambda_\gamma=-\frac{\sum _{\mu \nu}\int_{FS} \frac{\mathit{d^2{\bf k'_\mu}}}{|v_F({\bf k'_\mu})|}\int_{FS} \frac{\mathit{d^2{\bf k_\nu}}}{|v_F({\bf k_\nu})|} g_\gamma({\bf k'_\mu}) \Gamma_{\mu \nu}^{\mathrm{P}}({\bf k,k'}) g_\gamma({\bf k_\nu})}{\sum_\mu \int_{FS} \frac{\mathit{d^2{\bf k_\mu}}}{|v_F({\bf k_\mu})|} g^2_\gamma({\bf k_\mu})}
\label{lambdaEq}
\end{equation}
where $\gamma$ numerates the different waves projected ($s$,$p$,$d$ or $f$) and $\mathrm{P}$ depends on the $\gamma$ symmetry. $\mathrm{P}$ could be $\mathrm{singlet}$ or $\mathrm{triplet}$. The gap equation has a solution when $\lambda_\gamma$ is 1. We increase the interaction parameters until the dominant wave solves the equation, for stronger interactions the gap is already opened in that channel.

\section{Phase Diagram}

\begin{figure}
     \begin{tabular}{cc}
    \includegraphics[width=80mm]{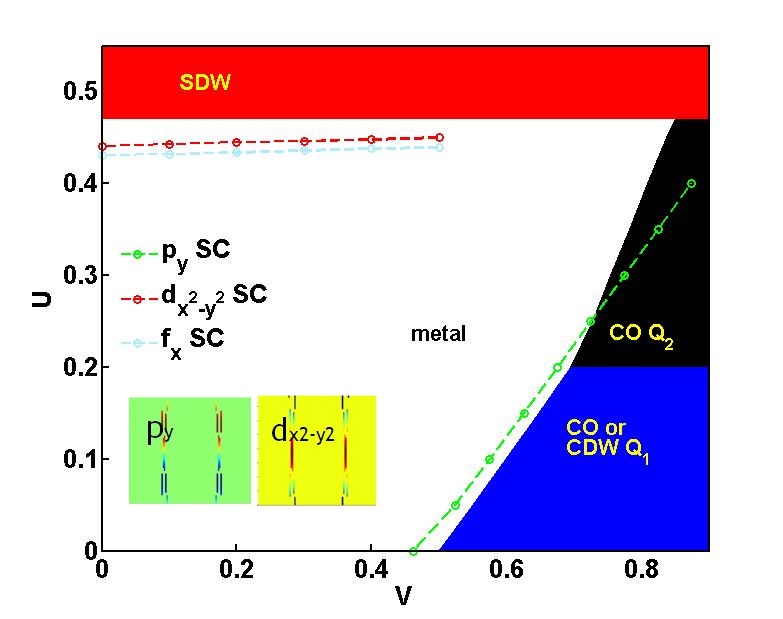}
  \end{tabular}
\caption{(Color online) 
(a) Phase Diagram $U-V$. Extended region of $p_y$ superconductivity with nodes in the Fermi surface (inset) close to the CO region. 
In the inset we show $d_{x^2-y^2}$ and $p_y$ wave functions.}
\label{PD}
\end{figure}

\begin{figure}
  \includegraphics[width=42mm]{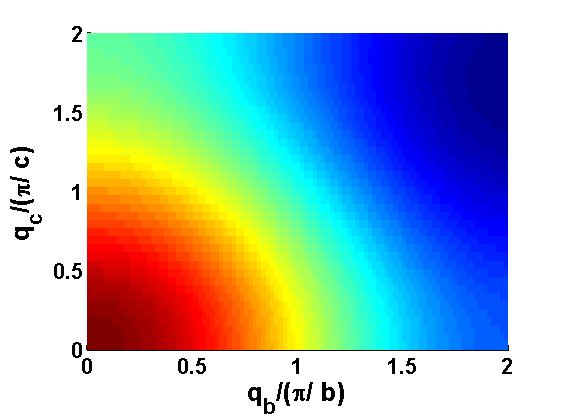}
  \includegraphics[width=42mm]{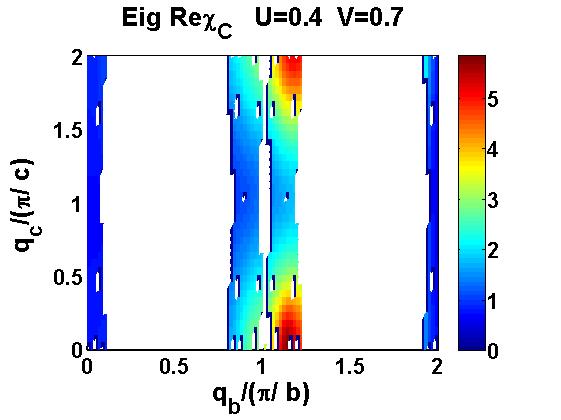}\\
\caption{(Color online) Left: momentum space distribution of the interaction, we show the sum of all components. Right: The same figure as Fig. \ref{ChargeOrderEig} (bottom right),  in those momentum relevant to the pairing vertex.}
\label{ChargeOrder}
\end{figure}

\begin{figure}
  \includegraphics[width=42mm]{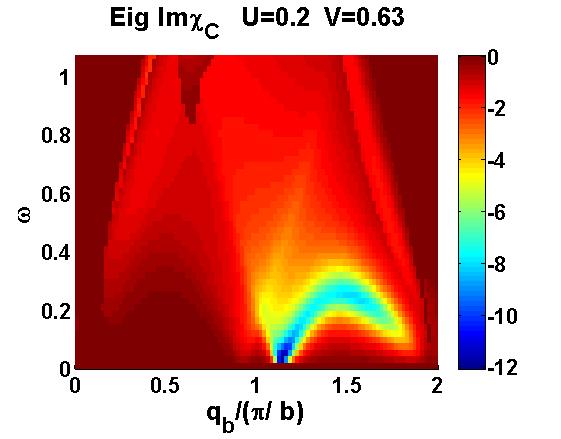}
  \includegraphics[width=42mm]{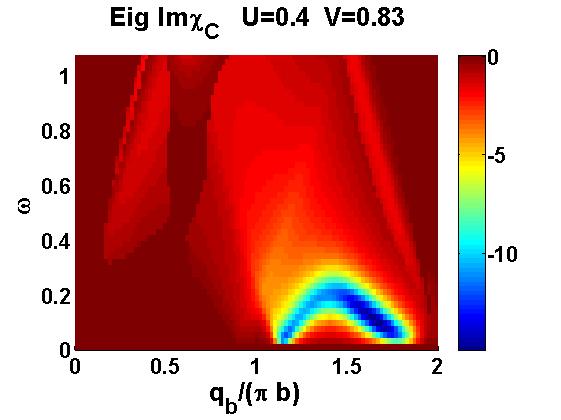}\\
  \includegraphics[width=42mm]{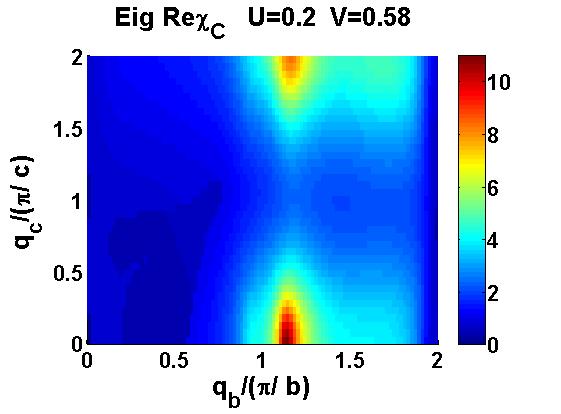}
  \includegraphics[width=42mm]{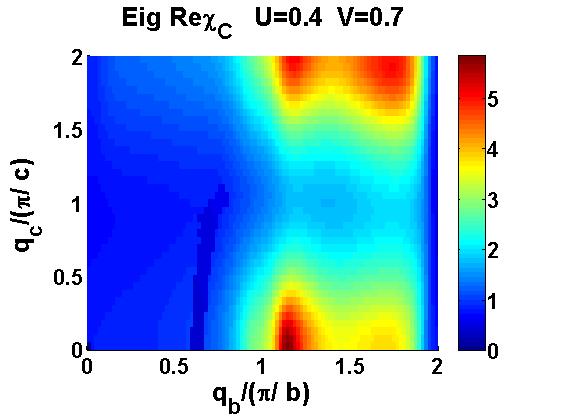}\\
\caption{(Color online) Top. Imaginary part of the larger eigenvalue of the charge susceptibility near the critical value (See Fig. \ref{PD}) for $U=0.2$ and ($U=0.3$) in left and (right) panel. Bottom. Real part of the larger eigenvalue in momentum space and zero frequency close to the critical value for $U=0.2$ and ($U=0.3$) in left and (right) panel.}
\label{ChargeOrderEig}
\end{figure}

\begin{figure}
     \begin{tabular}{cc}
    \includegraphics[width=80mm]{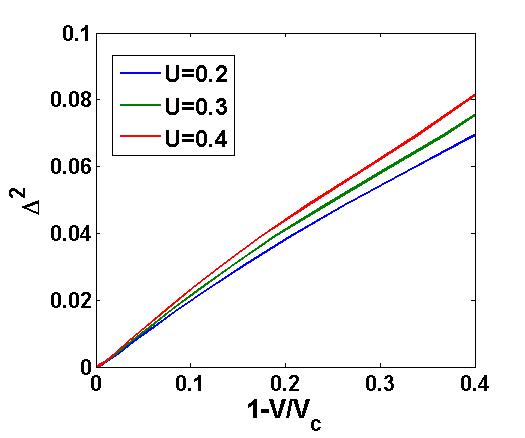}
  \end{tabular}
\caption{(Color online) 
Gap squared of the ${\bf Q_2}$ critical mode scaled with the critical interaction $V=V_c$. We observe a $\frac{1}{2}$ exponent near the critical value.}
\label{CMQ2}
\end{figure}

Using the parametraization described in section \ref{sec:model} we can study the complete parameter space, reduced to two variables $U$ and $V$. 
The RPA spin susceptibility (Eq. \ref{eq:chiS}) breaks at $U=0.47$ indicating a Spin Density Wave (SDW) phase.

The RPA charge susceptibility (Eq. \ref{eq:chiC}) diverges for different momenta for different $U$ on-site Hubbard interaction, leading to different charge order regions in the phase diagram (See Fig. \ref{PD}). 
The charge order susceptibility divergence consists in an interplay between the bare susceptibility strongly peaked at $q_b\approx \pi$ (Fig. \ref{chi0}(b)) and the charge interaction $U_c(\bf{q})$. The analysis involves $4 \times 4$ terms but in essence can be understood with the sum of the $16$ contributions.
 We observe that $U_c(\bf{q})$(Fig. \ref{ChargeOrder} c) is minimum at the $(2\pi,2\pi)$ edge of the Brillouin Zone, notice that the periodicity is not required since we are dealing with the sum of the elements of a matrix. For $U=0$, red means positive and blue negative, for that reason, among all nesting vectors ($q_b\approx \pi$), $\bf Q_1$ diverges first.  The divergence at this momentum stems from nesting.\\
As long as we increase $U$,  $U_c(\bf{q})$ remains negative in a smaller region, leading to the displacement of the divergence to $\bf Q_2$. Why $\bf Q_2$ does not change with $U$ can be understood from the bare susceptibility structure, $\chi_0$. $\chi_0$ in the entire Brillouin Zone (only $q_b$ matters) can be divided in three zones: $0<q_b \lesssim 0.1\pi/b$, $0.1\pi/b \lesssim q_b \lesssim 0.6\pi/b$ and $0.6\pi/b \lesssim q_b<\pi/b$ (and symmetric regions). In the first zone the susceptibility increases sharply due to the warping, increasing $q_b$ we can connect more points of the Fermi surface; in the second region the system only have access to the Fermi sheets at one side, increasing slowly the value of the particle-hole susceptibility; in the third region connections among the two pairs of sheets gives also a rapid enhancement. In our case, the range of $U$ below the SDW ordered phase, makes the negative $U_c(\bf{q})$ to be in the second region of the bare susceptibility, since this region is slowly q-dependent, we observe a minimal change of $\bf Q_2$ with increasing $U$. The divergence of the charge susceptibility at this momentum is due to interactions and the softening can be described as a critical mode similar to the one found in Ref. \onlinecite{JMandJV}. In Fig. \ref{CMQ2} we see a critical exponent of $\frac{1}{2}$. \\

The transition from ${\bf Q_1}$ to ${\bf Q_2}$ ordering phases, is also shown in Fig. \ref{ChargeOrderEig}. The upper panels show the frequency against momentum of the charge susceptibility (maximum eigenvalue, which is significantly larger than the other three), the lower panels show the charge susceptibility (maximum eigenvalue) at zero frequency. On the left hand panels $U=0.2$ while on the right panels $U=0.4$. We observe a change in the spectral weight of the collective mode from ${\bf Q_1}$ to ${\bf Q_2}$ when $U$ is increased. Moreover, while the weight at ${\bf Q_1}$ exists at any value of $V$, at ${\bf Q_2}$ the mode softens signaling at the proximity to the transition.\\

Near the SDW region we found superconductivity in $d_{x^2-y^2}$ channel.This behavior is consistent with the expected for a quasi-one dimensional square lattice at quarter filling \cite{japDwave}. A very recent preprint \cite{tripletLiPB}, proposes an order parameter with different sign in each band and a total of three node planes in the b-direction (and two more in the c-direction) at $V=0$. We skip the bracketed description and project the wave (here we call it $f_x$) with our methodology. The results Fig. \ref{PD} show that both $d$- and $f$-channels are very close, with the $f_x$ dominating. 


As long as experiments do not show signatures of SDW gap opening or magnetic response \cite{Matsuda}, we can work with lower $U$ values, to avoid strong spin fluctuations. The Coulomb interactions are comparable with \cite{nuss}.

\begin{figure}
     \begin{tabular}{cc}
    \includegraphics[width=40mm]{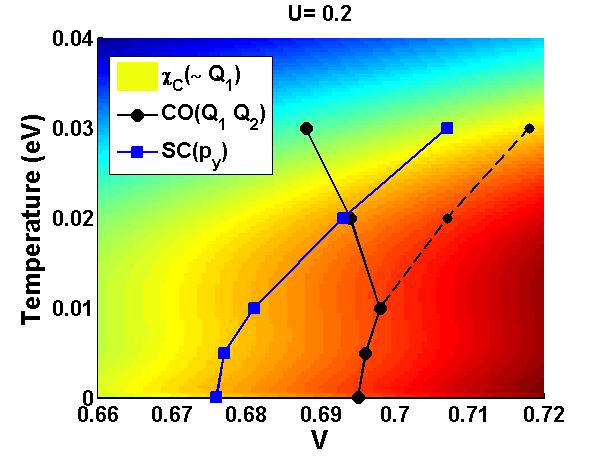}
    \includegraphics[width=40mm]{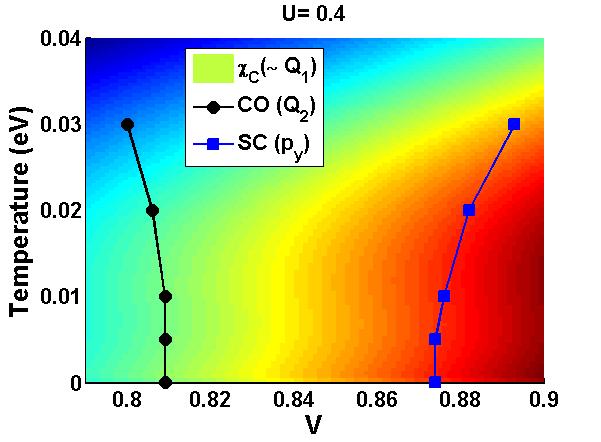}
  \end{tabular}
\caption{(Color online) 
The background represents the charge fluctuations near ${\bf Q_1}$ and on top of that are the CO transition line (green) and superconducting transition line (solid black) for $U=0.2$ and ($U=0.4$) in right (left) panel. The dashed black line is the CO transition due to $\bf Q_1$ if the other order were not present. The blue line is the superconducting transition.
}
\label{coexistencia}
\end{figure}

 Near the CDW or CO regions of the phase diagram we found triplet superconductivity in $p_y$ channel, with nodes at the Fermi surface.
 Near the $\bf Q_1$ CDW region, we found a narrow stripe \cite{GraserFootNote} of superconductivity due to charge fluctuations at $\bf Q_1$. We observe $\bf Q_1$ is a nesting vector connecting all the Fermi surface with different phase of the order parameter. See inset Fig. \ref{PD}\\
 From this study, apparently we can design the interactions Fig. \ref{ChargeOrder} (c) to be minimum in a given momentum, in such a way that favors superconductivity with a certain order parameter. Nevertheless, we need to take into account the bare susceptibility structure and the orbital distribution in real space. In the present model, the bare susceptibility is peaked at $q_b\approx\pi$, and the divergence at $\bf Q_1$ is favored by perpendicular Coulomb interactions $V_\perp$ and $W_\perp$ whereas interactions along the chains does not distinguish momenta in the $b$-direction.
 \\
 
 The $\bf Q_2$ momentum is not involved in the vertex calculation (Eq. \ref{lambdaEq}) so, we are still able to work with the superconducting vertex since it has not divergences. In that region, strong charge fluctuations still persist at $\bf Q_1$, due to nesting, and superconductivity would be found if another charge ordered phase were not present. 
\\
\subsection{Coexistence in the model}
If the order parameter of the CO is small, and assuming that the $\bf Q_2$ modulation does not open a gap at $E_F$,
we consider now the possibility of coexistence with SC in this model, even though it may not have relevance for the material.

We study the coexisting region with temperature (See Fig. \ref{coexistencia}). The charge fluctuations  have a reentrant behavior in RPA approach \cite{merino2006}, due to the fact that the bare susceptibility $\chi_0({\bf q},\omega)$ is maximum in energy ($\omega\ll t$), when ${\bf q}$ connects different points of the band structure approximately $\omega$ away from Fermi level. In that case $\bf Q_1$ is a nesting vector and the maximum is close in energy, $\omega=0.01\approx T$. However $\bf Q_2$ exhibits its peak of reentrant behavior at a larger energy, and we only see the decrease of critical $V$ with temperature. In Fig. \ref{coexistencia} (a) the critical momentum changes from $Q_1$ at low temperatures to $Q_2$. We observed that change from charge susceptibility and it is represented by the change of behavior of the CO line. The temperature makes the bare susceptibility softer, lowering the value of $\chi_0({\bf Q_1})$ and shifting the critical momentum to $Q_2$. 

\begin{figure}
     \begin{tabular}{cc}
   %
       \includegraphics[width=40mm]{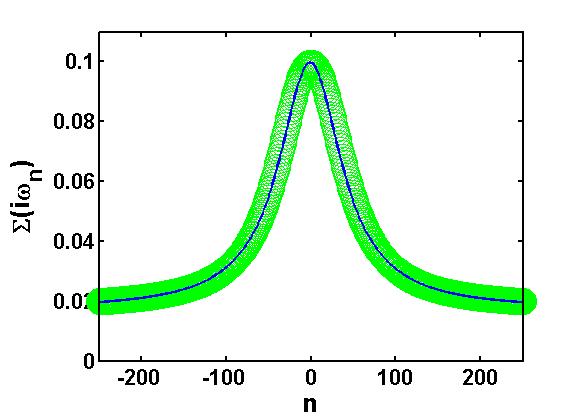}
   \includegraphics[width=40mm]{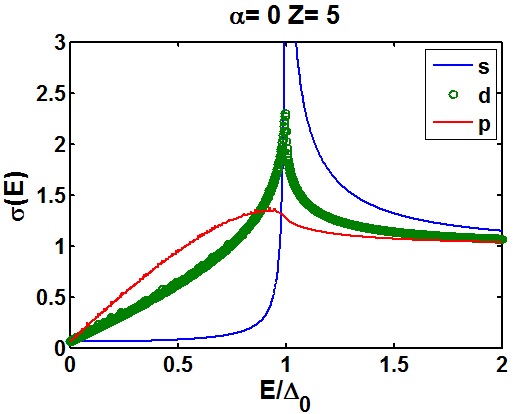}
  \end{tabular}
\caption{(Color online) 
Left.
Gap in Matsubara frequency ($i\omega_n$)at $T=0.01$, $U=0.2$ and $V=0.676$, green line. And Lorentizan fit, blue line. 
Right. Conductance against energy for different superconducting order parameters. We can distinguish $p_y$ wave. $\alpha$ is the angle of the order parameter with the junction. $Z$ is the height of the tunnel barrier, in this case the insulating phase. \cite{tanaka}
 }
\label{MVI}
\end{figure}

\section{Eliashberg equation with a reduced vertex}

In the previous section we have shown that the superconducting vertex is dominated by  the charge susceptibility near $\bf Q_1$,  see (Eq. (\ref{lambdaEq})). It results that using just a few vertex momenta we reproduce the $\lambda_{py}$ value. However, we cannot reduce easily the 4 orbital model for a simpler tight-binding, since near $\bf Q_1$ the bands has a similar weight in the four orbitals. For that reason, we select the larger values of the pairing vertex (calculated with the 4 orbital model) (Eq. \ref{pairingVertexs},\ref{pairingVertext}) and as a result, we see that less than 10\% of the vertex is enough to get more than 90\% of $\lambda_{py}$ value. In order to reproduce $\lambda_{py}$ from momenta near $\bf Q_1$ we need to multiply the value of the vertex by 4, otherwise we need to include momenta near ${\bf q}=0$ because it is significantly large,and connects many pairs of Fermi surface momenta.

 Moreover, we see that the value of the pairing vertex is almost independent of $q_y$. All those simplifications, allow us to work with a simpler model and solve the linear Eliashberg equation in Matsubara frequencies, given by:
 \begin{widetext}
 \begin{equation}
 \lambda_ {py} \Sigma({\bf k},i\omega_n)=\frac{-1}{N}
 \sum_{{\bf k'}i\omega_n'} G^0({\bf k'},i\omega_n')
  \Gamma^{\mathrm{triplet}}({\bf k,k'},i\omega_n-i\omega_n') G^0({\bf- k'},-i\omega_n') \Sigma({\bf k'},i\omega_n')
\end{equation}
\end{widetext}
 where 
 \begin{equation}
 G^0({\bf k'},i\omega_n')_{sp}=\sum_{\nu} \frac{a^s_\nu({\bf k}) a^p_\nu({\bf k})}{i\omega_n-\epsilon_\nu({\bf k}) )}
 \end{equation}
are 4 by 4 matrices, and $\Gamma^{\mathrm{triplet}}$ is also a matrix defined in Eq. (\ref{pairingVertext}) but with $ i\omega_n$ dependence coming from the bare susceptibility (Eq. \ref{eq:chi0} in Eq. \ref{GTriplet}). We calculate the momentum dependence of the gap, by projecting on the $p_y$ order parameter: $\Sigma({\bf k},i\omega_n)=f(i\omega_n)\sin(k_cc)$.
The result (shown in Fig. \ref{MVI}) can be fitted by a Lorentzian plus a constant; analytic continuation or Pad\'e approximants gives the same result, a real constant for the relevant frequencies. Provided the gap value is small, only low frequencies are relevant for experiments, as Normal-Insulator-Superconducting junctions \cite{tanaka}. (See Fig. \ref{MVI}(b))\\

\section{Discussion and Conclusions}

As was previously mentioned, Li$_{0.9}$Mo$_6$O$_{17}$ exhibits signatures of Luttinger liquid behavior for a wide range of temperatures. 
Thus, it is important to discuss its relation with the physics described above. 

\begin{figure}
     \begin{tabular}{cc}
   %
   \includegraphics[width=80mm]{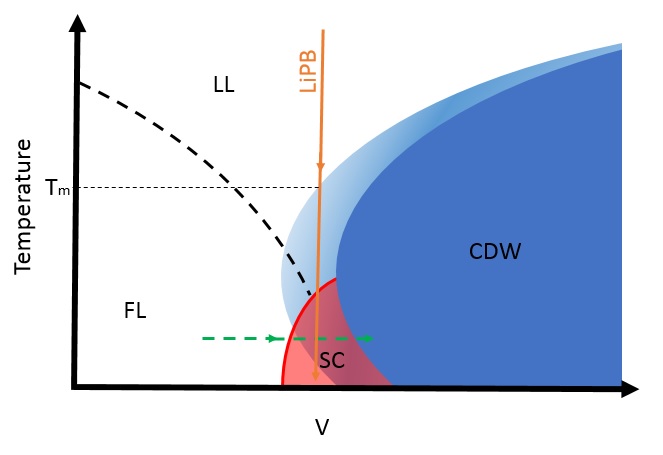}
  \end{tabular}
\caption{(Color online) 
Schematic phase diagram for the LiPB. The present study comprises the green horizontal arrow, we believe the temperature dependence of the real material is represented by the vertical orange arrow.
 }
\label{Diagrams}
\end{figure}

In Fig. \ref{Diagrams}, we present a schematic phase diagram for the model and consider its relevance for the physics of LiPB.  Merging renormalization group estimation of the crossover temperature, the RPA calculations for the  CDW and considering the fluctuation exchange in the superconducting vertex, we compose a schematic diagram Fig. \ref{Diagrams}. At high temperatures, the metallic phase is a LL. As the temperature goes down, the perpendicular hopping drives the system through a crossover to a Fermi liquid and the inter-chain Coulomb interactions through a thermodynamic phase transition to a CDW. Our  analysis of the SC vertex comprises the dashed horizontal line. Since we are working at temperatures well below $T_{LL}$, the use of RPA, as perturbation theory of the essentially free electron system, is well justified as a starting point. In other words, we are able to describe the superconductivity as an instability of a Fermi liquid, in spite of the normal phase of the material being a LL. On the other hand, the behavior of the material as the temperature goes down, seems to be represented by the solid vertical line. This statement is based on the spectroscopies \cite{cazalilla2005, dudy2013} at temperatures right above $T_c$.  The density of states show power-law behavior very similar to the ones observed at much higher temperatures and similar values of $\alpha$.   Placing the material slightly on the left of that vertical line would imply an interesting crossover from one NFL (the LL) to another NFL (FL +  strong charge fluctuations). At a purely qualitative level, no evidence of a Fermi edge developing at low temperatures has been observed and  the experimental values of $\alpha$ seem to increase (instead of decrease). However, both alternatives rely on details of the model and should be quantitatively contrasted with  the spectroscopies. 

\begin{figure}
    \includegraphics[width=80mm]{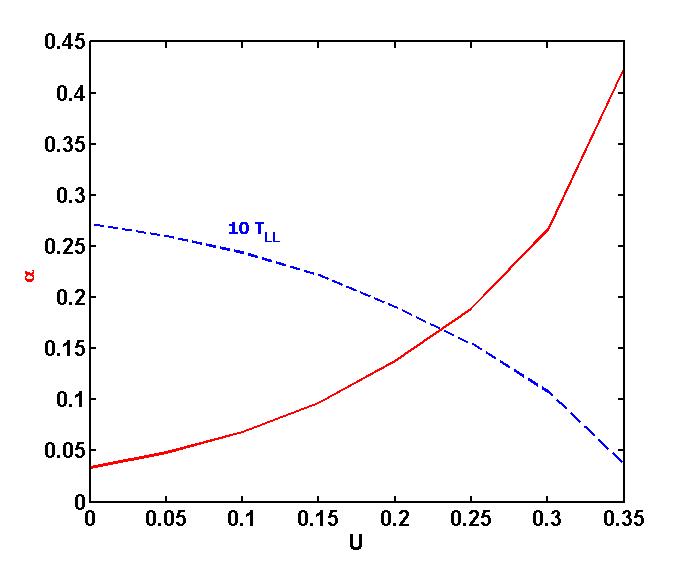}
\caption{(Color online) 
Dashed. Renormalization group estimation of the crossover temperature for the critical $V$ values found for each $U$. Solid. The exponent in the Luttinger density of states, $\alpha$.}
\label{TLL}
\end{figure}

The dashed line in Fig. \ref{Diagrams} shows the dimensional crossover from Luttinger liquid to Fermi Liquid \cite{Bois,GiamarchiChem}. The small value of the perpendicular hopping suggests considering it  as a perturbation.  Based on the renormalization group approach, we can estimate the crossover temperature to be  $T_{LL} \sim t\left(\frac{t_\perp}{t}\right)^{\frac{1}{1-\alpha}}$ where $\alpha$ is the exponent for the single-particle density of states. In Fig. \ref{TLL} we show the estimated dimensional crossover 

 for Luttinger chains coupled with Hubbard and V interactions. The value of $\alpha$ is computed using interaction parameters U and V following the CO border shown in Fig. \ref{PD}. W is set to zero. Note that the same Coulomb interactions driving the charge ordering,  allow for large values of $\alpha$.  Therefore, we expect $T_{LL}$ to be very small when the CDW is approached . This fact opens the possibility for a direct transition from the LL to the superconduting phase. It would be interesting to study this possibility with techniques similar to those used in Ref. \onlinecite{AlvarezCNT}. 

The charge ordering transition for RPA apparently occurs at arbitrarily large temperatures as V increases, but we expect the slight modifications  presented in  Ref. \onlinecite{SelfConsistentRPA}  which considers how the fluctuations effects modify the Greens function self-consistently, evaluting also vertex corrections. Other details  like the  reentrant behavior for the charge ordering transition, typical of RPA calculations,  are unessential for the physics of the system. 

To summarize, we have studied a microscopic extended Hubbard model for LiPB. We have characterized the couplings promoting SC close to different charge ordering patterns. A detailed analyisis within the RPA approximation of the vertex shows triplet superconductivity with nodes on the Fermi surface 
close to those ordered phases. The relevance of these results is discussed in terms of the general experimental perspective of the material.

\section*{Acknowledgments.} 
We thank J.W. Allen, J. Merino, L. Taillefer for fruitful discussions.
We acknowledge financial support from  MINECO FIS2012-37549-C05-03.

\end{document}